\documentclass{article}

\usepackage{arxiv}
\usepackage{subfig}
\usepackage{multirow}
\usepackage[utf8]{inputenc} % allow utf-8 input
\usepackage[T1]{fontenc}    % use 8-bit T1 fonts
\usepackage{hyperref}       % hyperlinks
\usepackage{url}            % simple URL typesetting
\usepackage{booktabs}       % professional-quality tables
\usepackage{amsfonts}       % blackboard math symbols
\usepackage{nicefrac}       % compact symbols for 1/2, etc.
\usepackage{microtype}      % microtypography
\usepackage{lipsum}
\usepackage{graphicx}
\graphicspath{ {./images/} }

\title{DePS: An improved deep learning model for de novo peptide sequencing}

\author{
 Cheng Ge \\
  Institute of Bioinformatics and Medical Engineering\\
  School of Electrical and Information Engineering\\
  Jiangsu University of Technology\\
  China, Changzhou 213001 \\
  \texttt{13851520957@163.com} \\
  \And
 Yi Lu \\
  Institute of Bioinformatics and Medical Engineering\\
  School of Electrical and Information Engineering\\
  Jiangsu University of Technology\\
  China, Changzhou 213001 \\
  \And
 Jia Qu \\
  School of Computer Science and Artificial Intelligence \& Aliyun School of Big Data\\
  Changzhou University\\
  China, Changzhou 213164 \\
  \And
 Liangxu Xie \\
  Institute of Bioinformatics and Medical Engineering\\
  School of Electrical and Information Engineering\\
  Jiangsu University of Technology\\
  China, Changzhou 213001 \\
  \And
 Feng  Wang \\
  School of Computer Science and Artificial Intelligence \& Aliyun School of Big Data\\
  Changzhou University\\
  China, Changzhou 213164 \\
  \And
 Hong Zhang \\
  School of Mathematics\\
  China University of Mining and Technology\\
  China, Xuzhou 221116 \\
  \And
 Ren Kong \\
  Institute of Bioinformatics and Medical Engineering\\
  School of Electrical and Information Engineering\\
  Jiangsu University of Technology\\
  China, Changzhou 213001 \\
  \texttt{  rkong@jsut.edu.cn} \\
  \And
 Shan Chang \\
  Institute of Bioinformatics and Medical Engineering\\
  School of Electrical and Information Engineering\\
  Jiangsu University of Technology\\
  China, Changzhou 213001 \\
  \texttt{schang@jsut.edu.cn} \\
}

\begin{document}
\maketitle
\begin{abstract}
De novo peptide sequencing from mass spectrometry data is an important method for protein identiﬁcation. Recently, various deep learning approaches were applied for de novo peptide sequencing and DeepNovoV2 is one of the represetative models. In this study, we proposed an enhanced model, DePS, which can improve the accuracy of de novo peptide sequencing even with missing signal peaks or large number of noisy peaks in tandem mass spectrometry data. It is showed that, for the same test set of DeepNovoV2, the DePS model achieved excellent results of 74.22\%, 74.21\% and 41.68\% for amino acid recall, amino acid precision and peptide recall respectively. Furthermore, the results suggested that  DePS outperforms DeepNovoV2 on the cross-species dataset. 
\end{abstract}

% keywords can be removed
%\keywords{First keyword \and Second keyword \and More}

\section{Introduction}
Protein identification is an essential step in proteomics research, and tandem mass spectrometry (MS/MS) is one of the principal methods to resolve this problem. Shotgun proteomics is a method to identify complex protein mixtures using a combination of high-performance liquid chromatography and mass spectrometry \cite{huang2012protein}. During the identification process, proteins are enzymatically digested ﬁrstly, and the resulting peptide mixtures are ionized and then scanned by MS/MS to obtain a set of spectra data. Finally, peptide identification and sequence assembly steps are performed to detect the amino acid sequence of the protein \cite{webb2007current}. In particular, mass spectrometry-based peptide identification is an important part of protein identification. Currently, there are two major categories for mass spectrometry-based peptide identification methods: (1) database searching, (2) de novo sequencing.
Database searching methods have been the main methods of peptide identification for decades. It compares the mass spectrometry data of the unknown peptide with the database data, looks for sequences that match the known peptide and selects the corresponding peptide with the highest match score. Common database search algorithms include Mascot \cite{perkins1999probability}, X!Tandem \cite{craig2004tandem}, pFind \cite{li2005pfind} and PEAKS DB \cite{zhang2012peaks}. However, the database search is only suitable for peptides already collected in the database. Once the query peptide does not exist in the database, then the method will no longer be suitable. Therefore, De novo peptide sequencing is a good solution as an alternative to database search methods. 
For de novo peptide sequencing, the amino acid of peptide could be obtained by using the mass difference between two fragment ions without database search. There are several algorithms for de novo peptide sequencing. Traditional algorithms include method of exhaustion and graph-based method. The exhaustive method \cite{sakurai1984paas,olson2006novo} lists all possible candidate peptides based on the mass of the parent ion, and then compares the candidate peptides with the experimental profile to find the best matching candidate peptide. This method is capable of identifying shorter peptide sequences, but is not suitable for longer peptide sequences or for data produced by low precision mass spectrometers. The graph-based method \cite{frank2005pepnovo,ma2003peaks,fischer2005novohmm,mo2007msnovo,chi2010pnovo,ma2015novor}, which is more commonly used, constructs a mass spectrum peak linkage graph. If the mass difference between two peaks is equal to the mass of an amino acid residue, the two mass spectrum peaks are added to the (V, E) graph as two vertices and an edge. Once the linkage map has been constructed, all possible pathways are searched to obtain candidate peptides. Finally, the candidate peptides are sorted and output by means of a scoring function. The disadvantage of this method is that it is not applicable for the data with lost ion peaks, which is occasionally produced by low accuracy mass spectrometer. Meanwhile, higher precision means that less nodes are merged and the generated spectrum graph has more vertices, which directly leads to a higher computational complexity. Muth et al. \cite{muth2018evaluating} provided a detailed overview of traditional algorithms and evaluated their performance on selected experimental and simulated MS/MS datasets. Similarly, Mohammed et al. \cite{mohammed2018visualizing} presented a theoretical framework and a data processing workflow for visualizing and comparing the results of these different types of algorithms. 
Deep learning has evolved rapidly over the past several decades. It has excelled in feature extraction in large and complex datasets and is widely used in proteomics research \cite{cai2020itp,shi2021deep}. Deep learning based de novo peptide sequencing algorithms have been developed in recent years, including DeepNovo \cite{tran2017novo}, pDeep \cite{zhou2017pdeep} and DeepNovoV2 \cite{qiao2019DeepNovoV2}.  Among them, DeepNovoV2 is a state-of-the-art model of deep learning-based approaches for de novo peptide sequencing due to its excellent feature extraction network and unique feature representation. Its model structure consists of a Long Short Term Memory (LSTM) network and a T-net model from the Point Net \cite{qi2017pointnet}.
However, de novo peptide sequencing methods are strongly influenced by the quality of the mass spectrometry data. To improve the accuracy of de novo sequencing in the low accuracy tandem mass spectrometry data, we proposed the DePS model. Specifically, for the problem of missing signal peaks, DePS model expands the feature dimension based on DeepNovoV2. Meanwhile, to address the problem of a large number of noisy peaks in tandem mass spectrometry data, we were inspired by the Deep Residual Shrinkage Networks \cite{zhao2019deep}, which performs well on signal data with noise, and constructed a CNN network with an attention mechanism.

\begin{figure*}[!htbp]
\centering
\includegraphics[scale=0.6]{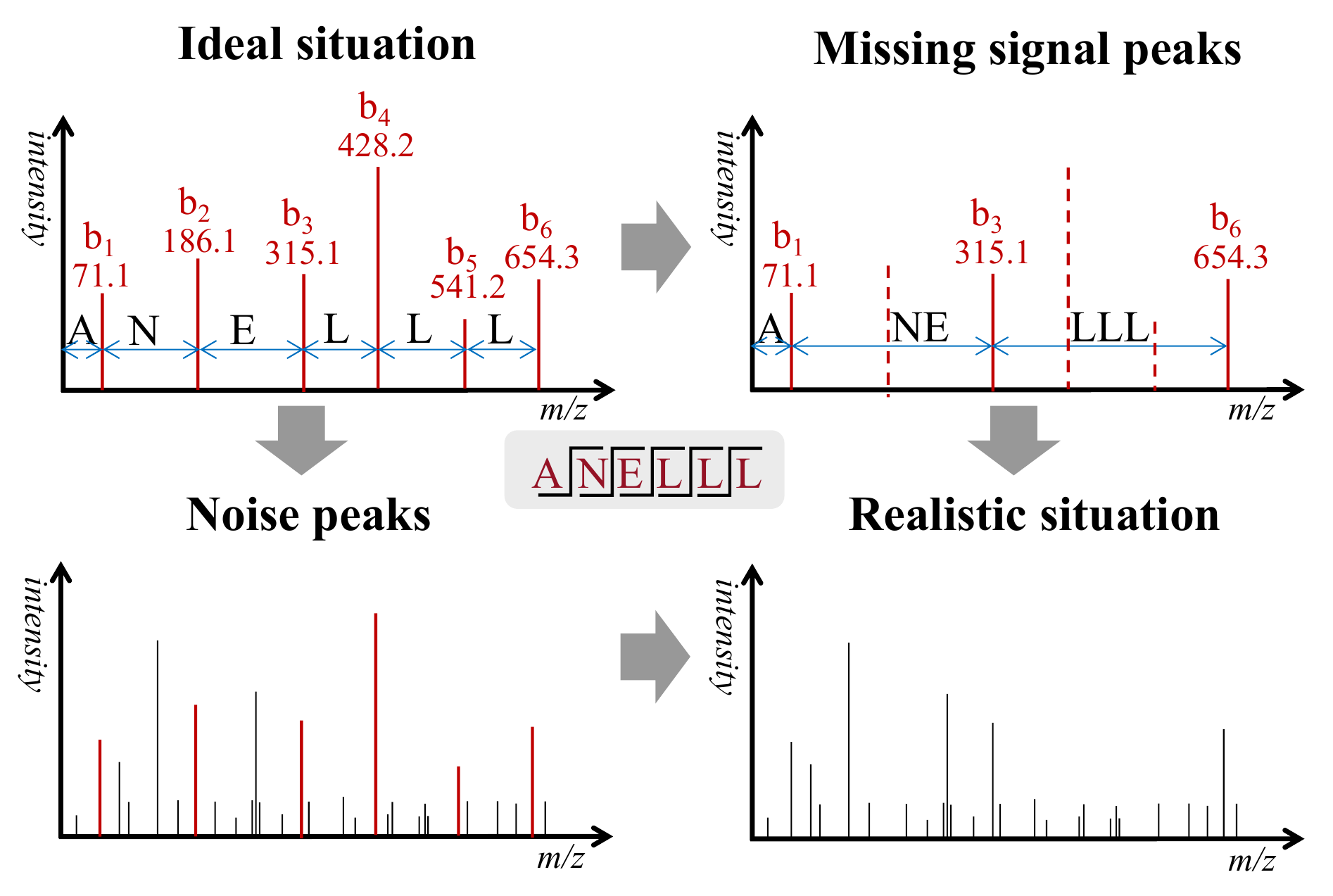}
\caption{Tandem mass spectrometry spectrum in different situations. Take the peptide sequence ”ANELLL” as an example, and only b-ions are shown here.  Under the ideal conditions,  the tandem mass spectrometry spectrum contains complete b-ion peaks.  Each neighbouring b-ion peak differs by one amino acid mass from each other.  In the case of missing b-ion peaks, the difference between neighbouring b-ion peaks is the mass of two or more amino acids.
In the case of including noise peaks, the b-ion peak is difficult to identify.  In realistic situations, they may even occur at the same time.}
\label{fig:1}
\end{figure*}

\section{Methods}
\subsection{Mass spectrometry data}
Mass spectrometry data includes a mass spectrometry and tandem mass spectrometry. The tandem mass spectrometry is obtained by fragmenting the peptides in a mass spectrometry. The tandem mass spectrometry mainly includes three complementary types of fragment ions, (a\_ion, x\_ion), (b\_ion, y\_ion) and (c\_ion, z\_ion), depending on different fragmentation methods and locations of the peptide chain break \cite{roepstorff1984proposal}. The most commonly used fragmentation methods are collision-induced dissociation (CID) and electron transfer dissociation (ETD). CID can produce b and y ions, and ETD mainly produces c and z ions. There are other types of ions that could be generated because of the incomplete dissociation of the peptide fragment, such as b and y ions that have lost ammonia (b-$\rm NH_3$, y-$\rm NH_3$) or water (b-$\rm H_2O$, y-$\rm H_2O$). We can plot the tandem mass spectra using the mass/charge (m/z) and intensity values of the ions. Normally, tandem mass spectrometry data may have some signal peaks lost or contain a large number of noise peaks, as shown in Figure~\ref{fig:1}, which adds difficulties to deep learning based de novo peptide sequencing algorithms. 

\begin{figure*}[!htbp]
\centering
\includegraphics[scale=0.6]{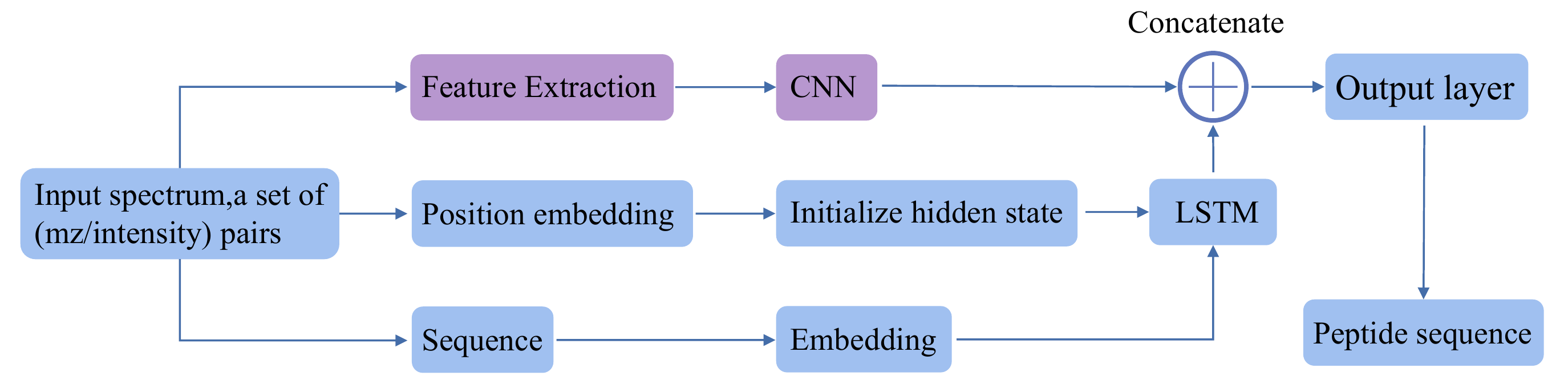}
\caption{The model structure of DePS. Firstly, different feature data is obtained from the mass spectrometry and these are passed into the CNN and LSTM respectively. After that, the output features of the two networks are concentrated and finally the peptide sequences are obtained through the output layer.}
\label{fig:2}
\end{figure*}

\subsection{Model construction}
The DEPS model mainly consists of feature extraction, CNN and LSTM module, as shown in Figure~\ref{fig:2}. It obtains feature representation of tandem mass spectrometry by using a feature extraction module and extracts features using a CNN module. The LSTM module is used to extract the language model information of peptides. It includes sequence features after embedding and hidden state that encoding by position embedding methods \cite{vaswani2017attention}.

\subsection{Feature Extraction}
\begin{figure*}[!htbp]
\centering
\includegraphics[scale=0.5]{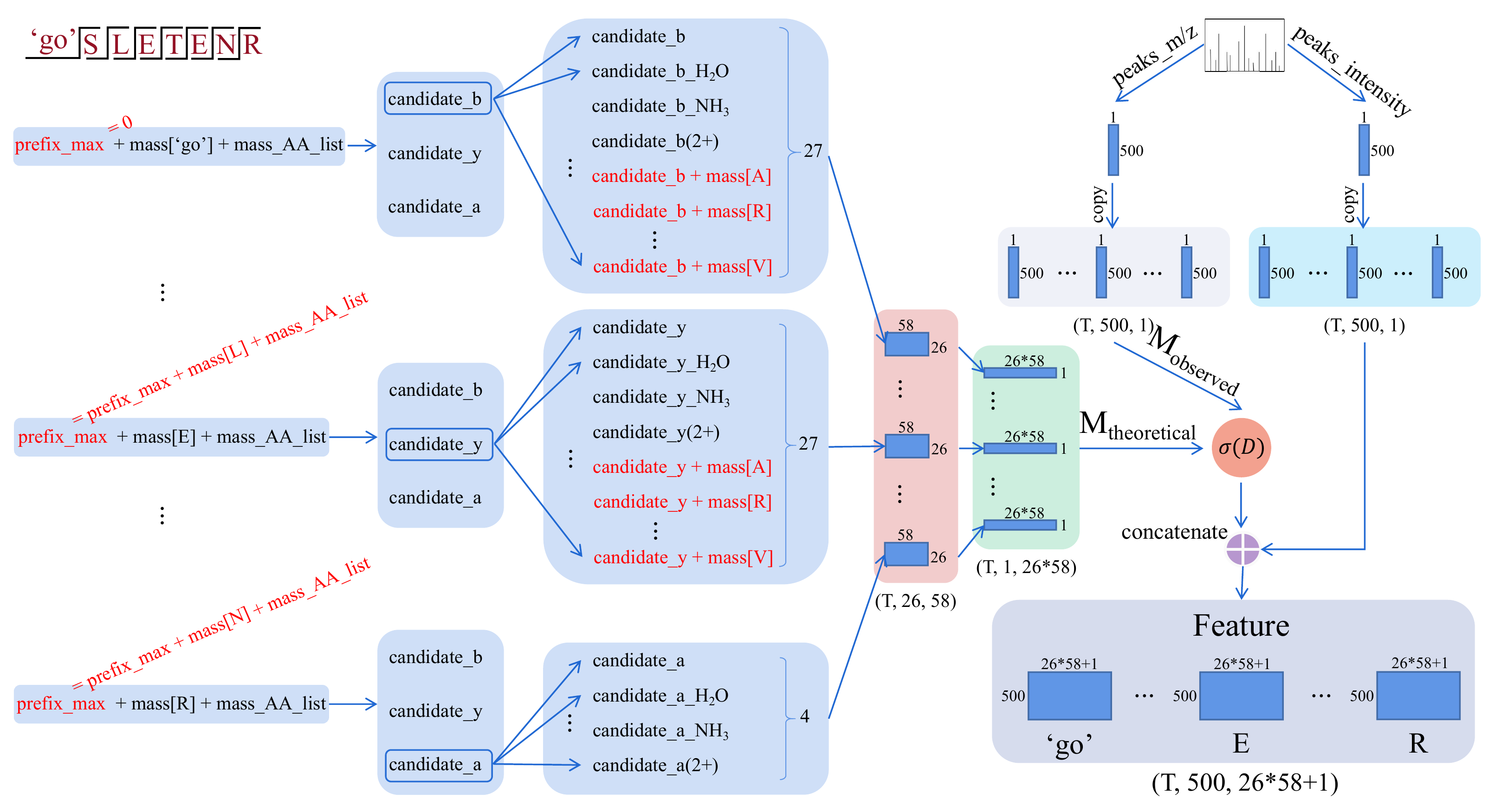}
\caption{The process of feature extraction. Where prefix\_max is an iterative variable with an initial value of 0, mass[E] represents the mass of amino acid E, mass\_AA\_list is a list of the masses of 26 tokens, and candidate\_b represents the mass of the candidate b-ion. T denotes the length of the sequence.}
\label{fig:3}
\end{figure*}

We use a set of [m/z, intensity] pairs to represent the tandem mass spectrometry spectrum and only select the top n most intense peaks (n = 500). A mass list, called mass\_AA\_list, is created. It contains  26  elements  (20  amino  acid  residues, 3 post-translational modifications,  and  3  special  tokens:  ”go”,  ”end” and  ”pad”). As shown in Figure~\ref{fig:3}, we used a feature extraction framework similar to DeepNovoV2. First, we get the theoretical ion masses in an iterative fashion based on the peptide sequence. It is a matrix of shape (v,k), which is denoted M$_{theoretical}$. After that, the observed ion masses are obtained from the spectrum, and matched to the theoretical ion masses. Finally, the obtained result is concatenated with the intensity values of the peaks to obtaining the feature matrix. Different from DeepNovoV2, we take into account the condition of signal peak loss to obtain the theoretical ion masses.

Take the peptide sequence "SLETENR" as an example. The sequence is preceded by an initial token ('go’, mass=0). After that, we use the mass of 'go' as the mass of the N-terminal, and add the resulting mass to  each  element  of  mass\_AA\_list. This step is used to obtain the theoretical mass of the b-ion (candidate b\_ion). Once the b-ion mass is obtained, the masses of y-ion and a-ion can be calculated using the mass relationship between the ions. Similarly, we can get candidate\_b\_H2O, candidate\_b\_NH3, candidate\_b(2+), candidate\_y\_H2O, candidate\_y(2+),
candidate\_y\_NH3, candida
te\_a\_H2O, candidate\_a(2+) and candidate\_a\_NH3 by using the mass relationship between the ions. We add 23 amino acid masses to each candidate ion mass individually, because, without considering the masses of the 3 special tokens masses, there are 23 possible arrangements of amino acids when the difference in mass between the signal peaks is two amino acid masses. Then, the feature of each token becomes (26,58) dimensions by dimensional transformation. This is a significant improvement compared to the feature dimension of DeepNovoV2 (26,12). This method is a good solution to the problem of missing key signal peaks. Further, in order to match with M$_{observed}$, the feature matrix dimension needs to be reconstructed as (1,26*58). Now, we have a matrix of M$_{theoretical}$. Next, the M$_{theoretical}$ matrix is matched with the M$_{observed}$ matrix obtained from the mass spectra to get the feature matrix by doing the operation of Equation (1). 

\begin{equation}
    \sigma(D)=\mathrm{e}^{\left(-\left|M_{observed}-M_{theoretical}\right| * c\right)}
\end{equation}
where c=100. Then, the peak intensities are concatenated with the feature matrix, as shown in Equation (2).\\
\begin{equation}
    F=\sigma(D) \oplus I
\end{equation}

In this way, the final feature matrix is obtained.

\subsection{CNN model}
Changes in the arrangement of amino acids in the peptide sequence will lead to the feature change of the tandem mass spectra. This constantly changing feature order makes it more difficult for CNNs to extract features. To solve this problem, DeepNovoV2 uses the T-net network \cite{qi2017pointnet}, which performs well in dealing with the problem of changing order of point cloud features. The T-net network consists of three 1D convolutions, one global-max-pooling, and two fully connected layers. The global-max-pooling operation which achieves order invariance makes the T-net network perform excellent in tasks with changing features.

\begin{figure*}[!htbp]
\centering
\includegraphics[scale=0.5]{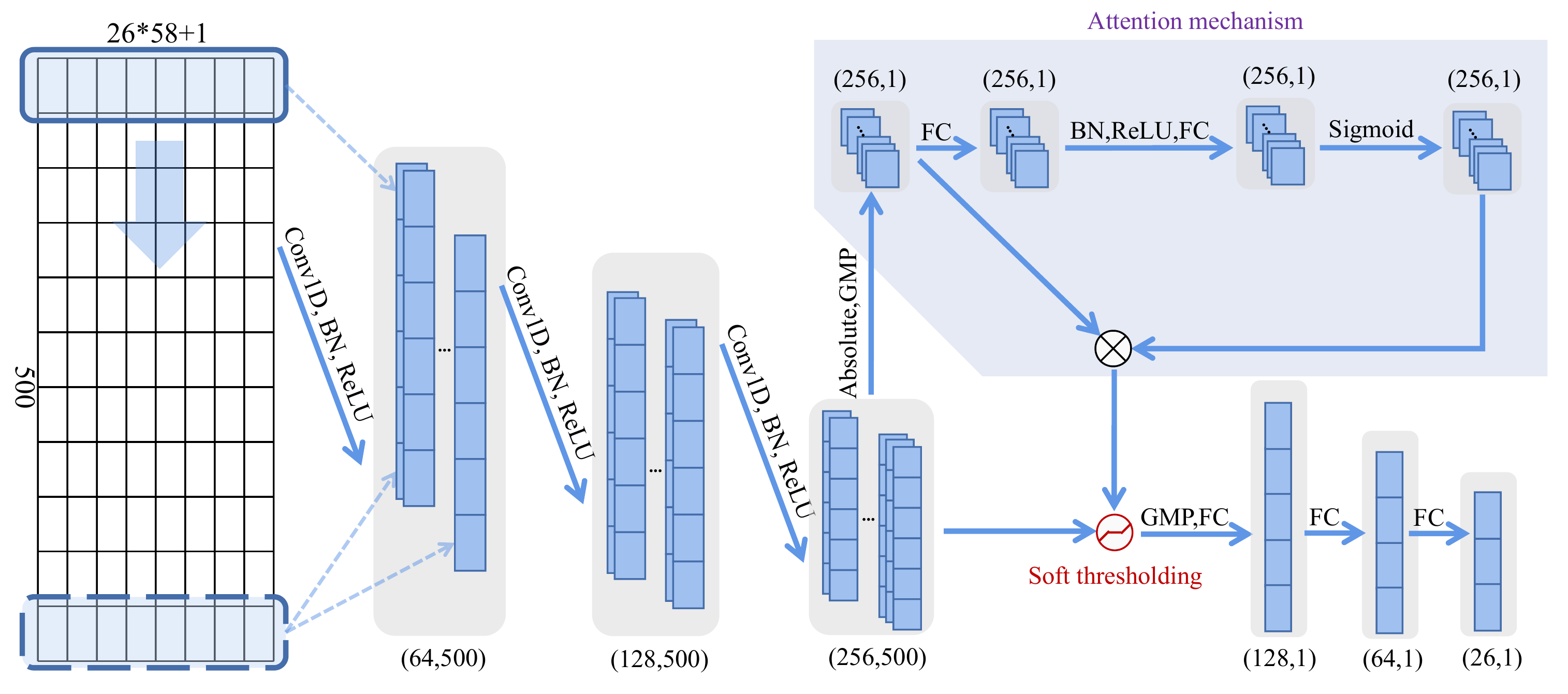}
\caption{DePS-net. It consists mainly of three one-dimensional convolutional layers, a deep shrinkage network module, a global max-pooling layer and three fully connected layers.}
\label{fig:4}
\end{figure*}

We propose a novel CNN model, which we call DePS-net, retains the global-max-pooling layer and adds a deep shrinkage network module, as shown in Figure~\ref{fig:4}. DePS-net consists mainly of three one-dimensional convolutional layers, a deep shrinkage network module \cite{zhao2019deep}, a global max-pooling layer and three fully connected layers.  The deep shrinkage network module consists mainly of attention mechanism and soft thresholding. The model adaptively thresholds through an attention mechanism, where each channel of the feature corresponds to a threshold. It can set separate thresholds for each feature on a feature-by-feature basis, and is therefore suitable for situations where the noise content within each feature varies. Soft thresholding is a function that shrinks the input data towards zero and is often used in signal noise reduction algorithms. It is formulated as follows:

\begin{equation}
y = \left\{\begin{array}{ll}
x-\tau & x>\tau \\
0 & -\tau \leq x \leq \tau \\
x+\tau & x<-\tau
\end{array}\right.
\end{equation}

where x denotes the input feature, y denotes the output feature and $\tau$ denotes the threshold value. The derivative of the soft thresholding function is formulated as follows:

\begin{equation}
\frac{\partial y}{\partial x}=\left\{\begin{array}{ll}
1 & x>\tau \\
0 & -\tau \leq x \leq \tau \\
1 & x<-\tau
\end{array}\right.
\end{equation}
We can see that the derivative of the soft thresholding function is either zero or one. 
Thus, unimportant features are eliminated through a combination of attention mechanism and soft thresholding.

In DePS-net model, features are first convolved by three 1D convolution layers (1D convolutional convolution, batch normalization (BN), ReLU activation). Then they are fed into a deep shrinkage network module. The specific steps of deep shrinkage network module include: (1) Changing the features into absolute values  and performing global-max-pooling; (2) Passing through a fully connected layer, a batch normalization layer, a ReLU activation layer, a fully connected layer, and a Sigmoid activation function layer in order; (3) Weighting the result with the result after the global-max-pooling layer; (4) Shorting the features with the results after 1D convolution layers and performing a soft thresholding operation. After that, the features are subjected to global-max-pooling. Finally, the features are passed through three fully connected layers.

\section{Experiments And Results}
\subsection{Dataset}
The same dataset used in DeepNovoV2 are used here, the DDA dataset of Hela samples\footnote{PRG 2018: Evaluation of Data-Independent Acquisition (DIA) for Protein Quantification in Academic and Core Facility Settings. \url{https://abrf.org/research-group/proteomics-research-group-prg}} (denoted as ABRF dataset). It is divided into training set, validation set and test set in the same way as in DeepNovoV2. As the ABRF dataset consists of only a single species, we also test DePS with cross species data published by Tran et al.\cite{tran2017novo} (MSV000081382), which contains 9 different species data.
\subsection{Implementation Details}
The implementation is based on Pytorch using an NVIDIA Tesla V100 32GB GPU. The model is trained using the Adam optimizer and a batch size of 32 with a learning rate of 0.001. We use focal loss as suggested by Tran et al. \cite{tran2017novo} when training the model.

To compare the performance of DePS with DeepNovoV2, we set up two tasks. Task 1 contains two subtasks. The CNN model is used in subtask 1, which is used to compare the performance of T-net with DePS-net in different feature dimensions. A CNN+LSTM structure is used in subtask 2, which is used to compare the performance of DeepNovoV2 with DePS in different feature dimensions.
Task2 is used to test the generalisation ability of DeepNovoV2 and DePS on a cross species dataset.

\subsection{Evaluation metrics}
In order to evaluate the performance of the model, we use the same evaluation metrics as those proposed by DeepNovoV2. The evaluation metrics include Amino Acid Recall (AAR), Amino Acid Precision (AAP), and Peptide Recall (PR).
\begin{equation}
    AAR=\frac{recall\_AA\_total}{target\_len\_total}
\end{equation}
\begin{equation}
    AAP =\frac{recall\_AA\_total}{predicted\_len\_total}
\end{equation}

\begin{equation}
    PR=\frac{recall\_peptide\_total}{predicted\_count}
\end{equation}
where recall\_AA\_total represents the total number of amino acids that were correctly predicted, target\_len\_total represents the total number of target amino acids, predicted\_len\_total represents the total number of predicted amino acids, recall\_peptide\_total represents the total number of peptides that were correctly predicted, and predicted\_count represents the total number of predicted peptides.
\subsection{Experiment Results}
\begin{figure}[!htbp]
\centering
\includegraphics[scale=0.4]{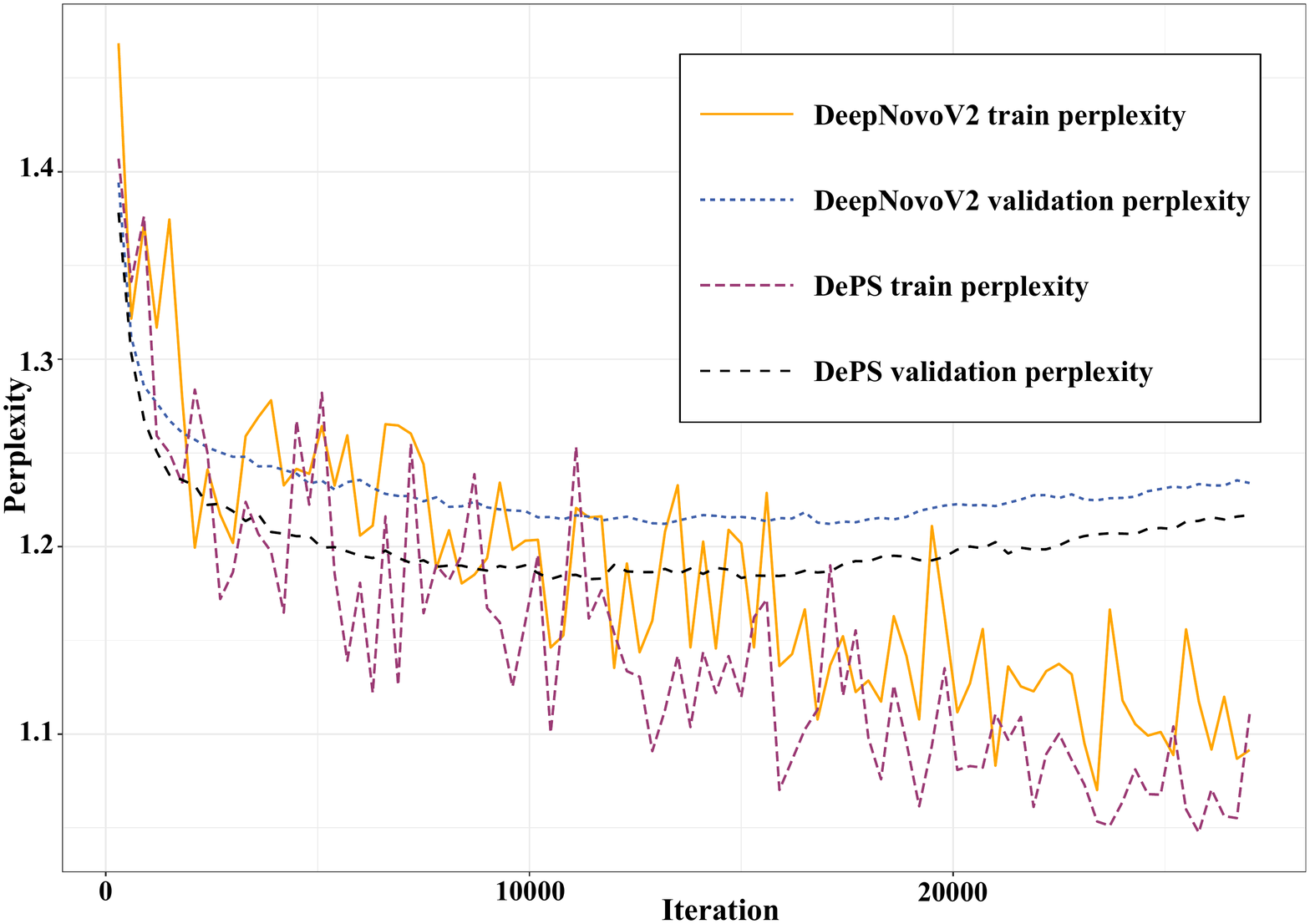}
\caption{The loss curves of DePS and DeepNovoV2 on the training set and validation set.
\label{fig:5}}
\end{figure}

\begin{figure}[!htbp]
\centering
\includegraphics[scale=0.4]{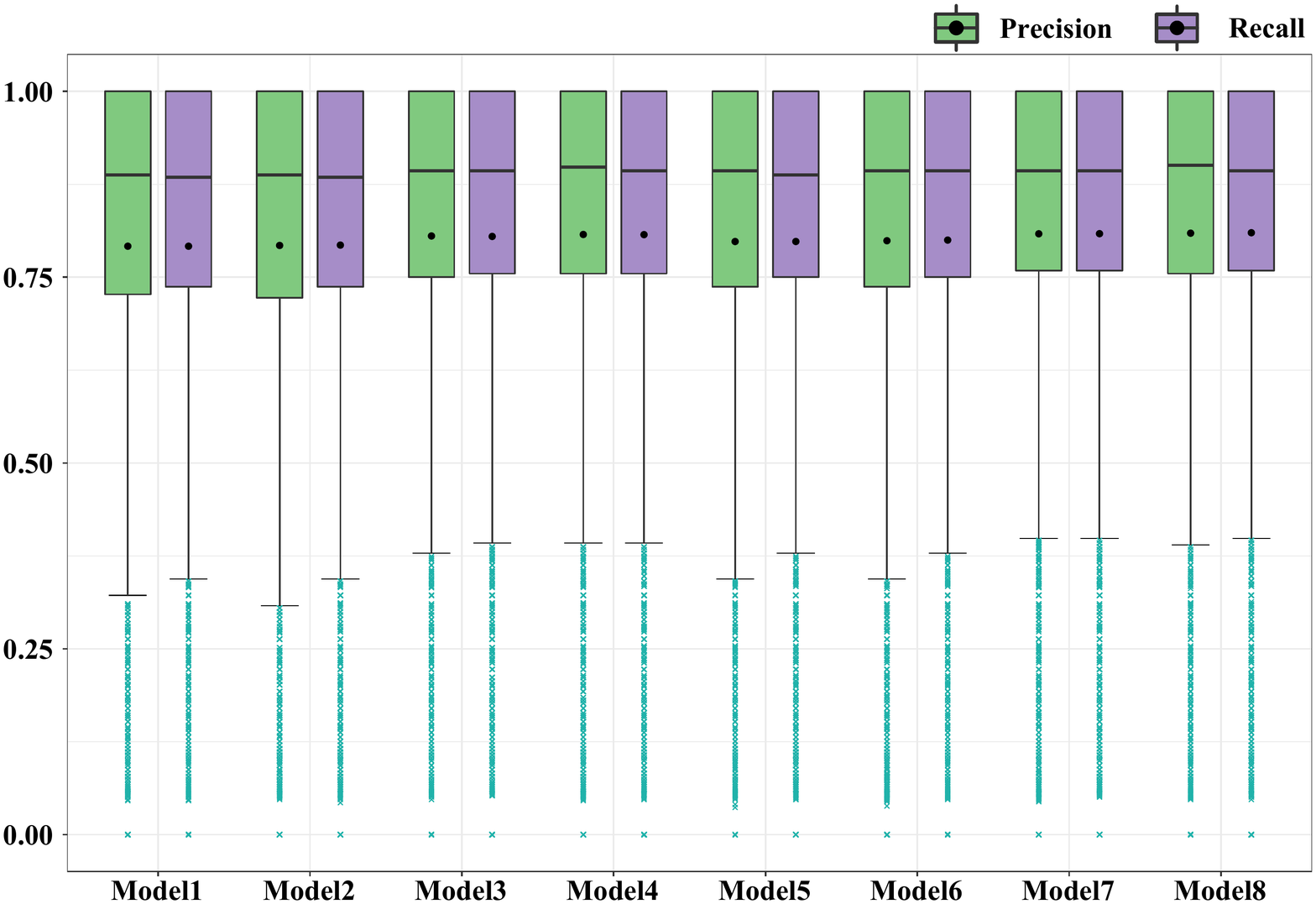}
\caption{Box plots of test results for the eight models. It shows the precision and recall of each peptide in the results of each model. \label{fig:6}}
\end{figure}

\textbf{Results of feature dimension expansion}\\
The experimental results of Task1 are shown in Table~\ref{tab1}. 
Comparing Model1 and Model3, Model2 and Model4, it can be seen that, with the model structure unchanged, it achieves a better result in all three evaluation metrics by extending the feature dimensions whether T-net or DePS-net. As the dimension expands from 12 to 58, DePS-net has increased its value in the AAR by 1.79 percent. In terms of AAP, this value is 1.95 percent. In terms of PR, it is 1.64 percent. Similarly, T-net has increased its value in the AAR by 1.77 percent. In terms of AAP, this value is 1.96 percent. In terms of PR, it is 1.26 percent. Hereby, we construct expanded features that include masses of original ions and the masses of original ions plus one extra amino acid. The expanded features allow CNN models to extract more useful information and could alleviate the problems of signal peaks missing.\\
\textbf{Results of different feature extraction models}\\
Comparing Model1 and Model2, Model3 and Model4, it can be seen that DePS-net outperforms T-net on all three evaluation metrics with the same feature dimensions. When the feature dimensions are 12, DePS-net is 0.19 percent higher than T-net in AAR. In terms of AAP, it is 0.12 percent higher than T-net. In terms of PR, it is 0.66 percent higher than T-net. When the feature dimensions are 58, DePS-net is 0.21 percent higher than T-net in AAR. In terms of AAP, it is 0.11 percent higher than T-net. In terms of PR, it is 1.04 percent higher than T-net. This suggests that the DePS-network model can 
slightly improve the accuracy of de novo peptide sequencing in the presence of a large number of noise peaks.\\
\textbf{Results of CNN+LSTM structure}\\
Comparing Model1 and Model5, Model2 and Model6, Model3 and Model7, Model4 and Model8, it can be seen that the overall performance of the models has been improved with the LSTM model added in the model.\\
\textbf{DePS vs DeepNovoV2}\\
Comparing the loss curves of DePS and DeepNovoV2 on the training set and validation set, as shown in Figure~\ref{fig:5}, it can be seen that DePS can converge faster and outperforms DeepNovoV2 as increasing itration. Comparing Model5 and Model8 in Table~\ref{tab1}, it can be seen that the results of DePS model show a significant improvement over DeepNovoV2. DePS is 1.49 percent higher than DeepNovoV2 in AAR. In terms of AAP, it is 1.47 percent higher than DeepNovoV2. In terms of PR, it is 1.87 percent higher than DeepNovoV2. Further, we analysed the precision and recall of each peptide, as shown in Figure~\ref{fig:6}, which clearly shows the performance of the different models in the test set. Comparing model 5 with model 8, it can be seen that the mean and lower bound of precision and recall for the peptides of DePS are higher than those of DeepNovoV2. 

\begin{table*}[!htbp]
\centering
\caption{Experimental results of Task 1.\label{tab1}}
\begin{tabular}{lllcccc}
\toprule
\textbf{Task}  &\textbf{Model}  & \textbf{Model structure}  & \textbf{Num of features}     & \textbf{AAR (\%)} & \textbf{AAP (\%)} & \textbf{PR (\%)}  \\ 
\toprule
\multirow{4}{*}{subtask1}  & Model1   & T-net & 12  & \underline{71.83}   & \underline{71.91}    & \underline{38.12}    \\
& Model2   & DePS-net & 12  & 72.02    & 72.03    & 38.78    \\
& Model3   & T-net & 58  & 73.60    & 73.87    & 39.38    \\
& Model4   & DePS-net & 58 & \textbf{73.81}    & \textbf{73.98}    & \textbf{40.42}    \\
\midrule
\multirow{4}{*}{subtask2}  &Model5 (DeepNovoV2)   & T-net+LSTM & 12 & \underline{72.73}   & \underline{72.74}    & \underline{39.81}    \\
& Model6   & DePS-net+LSTM & 12  & 73.00    & 72.80    & 40.52    \\
& Model7   & T-net+LSTM & 58  & 74.16    & 74.18    & 40.59    \\
& Model8 (DePS)   & DePS-net+LSTM & 58  & \textbf{74.22}    & \textbf{74.21}    & \textbf{41.68}    \\
\bottomrule
\end{tabular}
\end{table*}

\begin{figure*}[!t]
\centering
    	\subfloat{\includegraphics[scale=0.62]{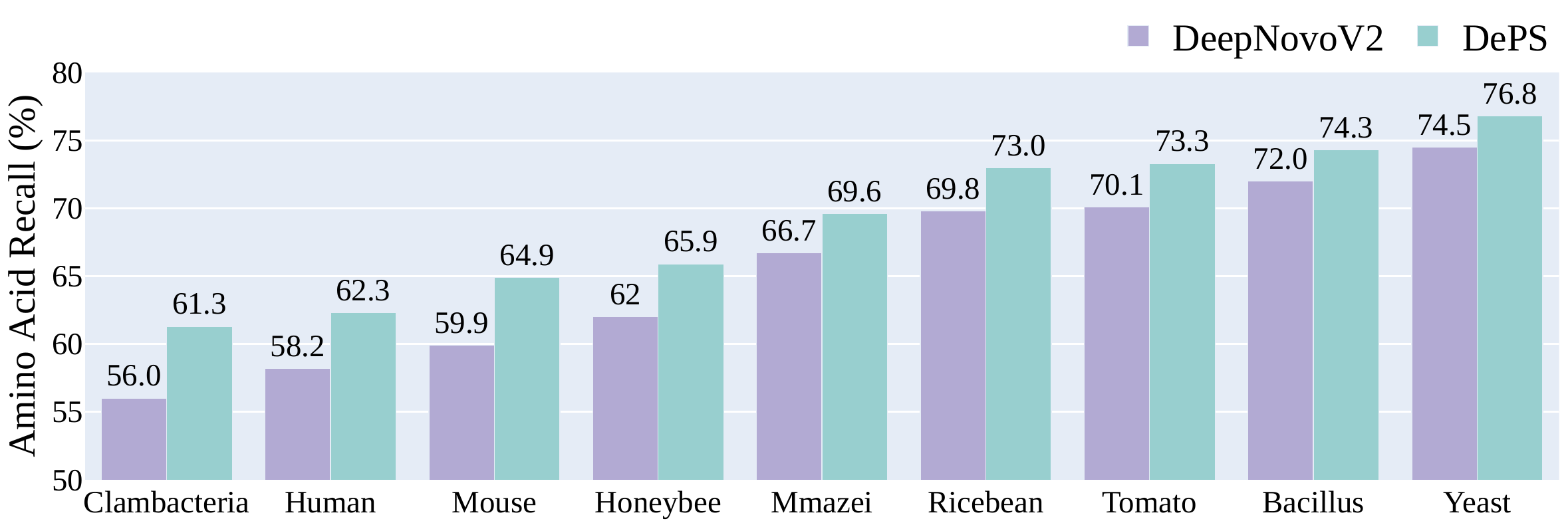}}\\
    	\subfloat{\includegraphics[scale=0.62]{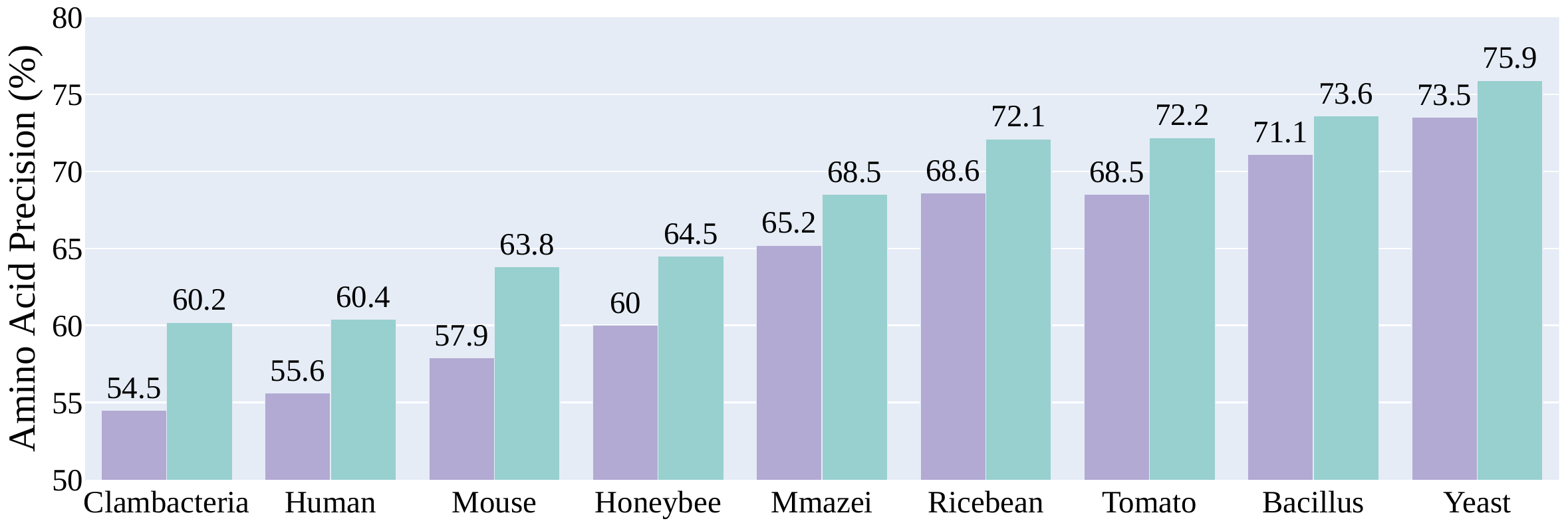}}\\
    	\subfloat{\includegraphics[scale=0.62]{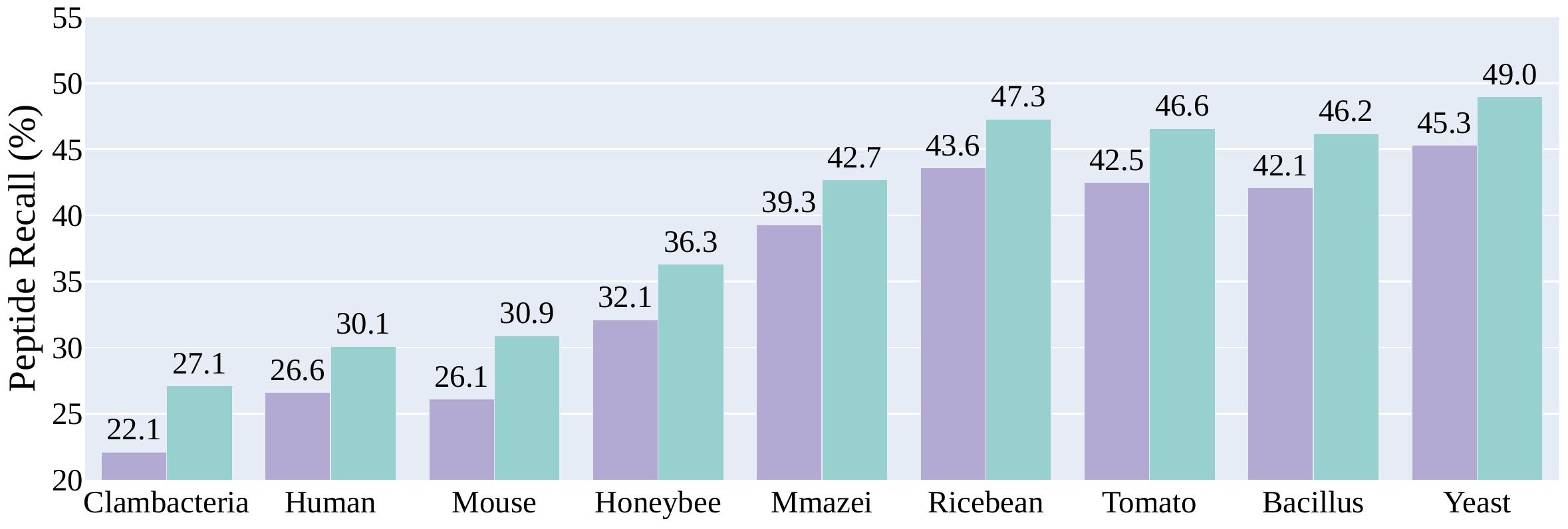}}\\
\caption{Bar Chart of experimental results for Task 2. Amino acid recall, amino acid precision and peptide recall of DeepNovoV2 and DePS.}\label{fig:7}
\end{figure*}  

To validate the generalization ability of DePS, we conduct experiments on cross species datasets. The experimental results of Task 2 are shown in figure~\ref{fig:7}, and it can be seen that DePS still performs well on cross species datasets. DePS is 2.3 to 5.3 percent higher than DeepNovoV2 in Amino Acid Recall. In terms of Amino Acid Precision, it is 2.4 to 5.7 percent higher than DeepNovoV2. In terms of Peptide Recall, it is 3.4 to 5 percent higher than DeepNovoV2. Moreover, DePS outperforms DeepNovoV2 in all three evaluation metrics on cross species datasets. In particular, on datasets where DeepNovoV2 predicts poorly, DePS has a large improvement. This situation occurs because  the  loss  of  signal  peaks  and a large number of noisy peaks can  lead  to  poor  predictions from DeepNovoV2,  whereas DePS is better able to solve these problems.

\section{Conclusion and Future Research}
This paper proposed a deep learning-based model for de novo peptide sequencing, called DePS. It achieves better performance than DeepNovoV2 by increasing the feature dimension and constructing a novel CNN network. By expanding the dimensionality of the feature matrix, DePS solves the problem of missing key signal peaks in tandem mass spectra very well. By proposing a DePS-net model with global-max-pooling, an attention mechanism and soft thresholding, DePS effectively improves the accuracy of de novo peptide sequencing in the presence of a large number of noise peaks. On ABRF dataset, the DePS model achieved excellent results of 74.22\%, 74.21\% and 41.68\% for amino acid recall, amino acid precision and peptide recall respectively. This result is significantly better than DeepNovoV2. On additional cross species data, the DePS model outperformed DeepNovoV2, which also shows that our model has great generalization ability. In the future, we will optimise the feature representation and  improve the feature extraction network.

\newpage

\bibliographystyle{unsrt}  
\bibliography{reference}

\begin{thebibliography}{10}

\bibitem{huang2012protein}
Ting Huang, Jingjing Wang, Weichuan Yu, and Zengyou He.
\newblock Protein inference: a review.
\newblock {\em Briefings in bioinformatics}, 13(5):586--614, 2012.

\bibitem{webb2007current}
Bobbie-Jo~M Webb-Robertson and William~R Cannon.
\newblock Current trends in computational inference from mass
  spectrometry-based proteomics.
\newblock {\em Briefings in bioinformatics}, 8(5):304--317, 2007.

\bibitem{perkins1999probability}
David~N Perkins, Darryl~JC Pappin, David~M Creasy, and John~S Cottrell.
\newblock Probability-based protein identification by searching sequence
  databases using mass spectrometry data.
\newblock {\em ELECTROPHORESIS: An International Journal}, 20(18):3551--3567,
  1999.

\bibitem{craig2004tandem}
Robertson Craig and Ronald~C Beavis.
\newblock Tandem: matching proteins with tandem mass spectra.
\newblock {\em Bioinformatics}, 20(9):1466--1467, 2004.

\bibitem{li2005pfind}
Dequan Li, Yan Fu, Ruixiang Sun, Charles~X Ling, Yonggang Wei, Hu~Zhou, Rong
  Zeng, Qiang Yang, Simin He, and Wen Gao.
\newblock pfind: a novel database-searching software system for automated
  peptide and protein identification via tandem mass spectrometry.
\newblock {\em Bioinformatics}, 21(13):3049--3050, 2005.

\bibitem{zhang2012peaks}
Jing Zhang, Lei Xin, Baozhen Shan, Weiwu Chen, Mingjie Xie, Denis Yuen, Weiming
  Zhang, Zefeng Zhang, Gilles~A Lajoie, and Bin Ma.
\newblock Peaks db: de novo sequencing assisted database search for sensitive
  and accurate peptide identification.
\newblock {\em Molecular \& cellular proteomics}, 11(4):M111--010587, 2012.

\bibitem{sakurai1984paas}
Tsuneaki Sakurai, Takekiyo Matsuo, Hideo Matsuda, and Ituso Katakuse.
\newblock Paas 3: A computer program to determine probable sequence of peptides
  from mass spectrometric data.
\newblock {\em Biomedical mass spectrometry}, 11(8):396--399, 1984.

\bibitem{olson2006novo}
Matthew~T Olson, Jonathan~A Epstein, and Alfred~L Yergey.
\newblock De novo peptide sequencing using exhaustive enumeration of peptide
  composition.
\newblock {\em Journal of the American Society for Mass Spectrometry},
  17(8):1041--1049, 2006.

\bibitem{frank2005pepnovo}
Ari Frank and Pavel Pevzner.
\newblock Pepnovo: de novo peptide sequencing via probabilistic network
  modeling.
\newblock {\em Analytical chemistry}, 77(4):964--973, 2005.

\bibitem{ma2003peaks}
Bin Ma, Kaizhong Zhang, Christopher Hendrie, Chengzhi Liang, Ming Li, Amanda
  Doherty-Kirby, and Gilles Lajoie.
\newblock Peaks: powerful software for peptide de novo sequencing by tandem
  mass spectrometry.
\newblock {\em Rapid communications in mass spectrometry}, 17(20):2337--2342,
  2003.

\bibitem{fischer2005novohmm}
Bernd Fischer, Volker Roth, Franz Roos, Jonas Grossmann, Sacha Baginsky, Peter
  Widmayer, Wilhelm Gruissem, and Joachim~M Buhmann.
\newblock Novohmm: a hidden markov model for de novo peptide sequencing.
\newblock {\em Analytical chemistry}, 77(22):7265--7273, 2005.

\bibitem{mo2007msnovo}
Lijuan Mo, Debojyoti Dutta, Yunhu Wan, and Ting Chen.
\newblock Msnovo: a dynamic programming algorithm for de novo peptide
  sequencing via tandem mass spectrometry.
\newblock {\em Analytical chemistry}, 79(13):4870--4878, 2007.

\bibitem{chi2010pnovo}
Hao Chi, Rui-Xiang Sun, Bing Yang, Chun-Qing Song, Le-Heng Wang, Chao Liu, Yan
  Fu, Zuo-Fei Yuan, Hai-Peng Wang, Si-Min He, et~al.
\newblock pnovo: de novo peptide sequencing and identification using hcd
  spectra.
\newblock {\em Journal of proteome research}, 9(5):2713--2724, 2010.

\bibitem{ma2015novor}
Bin Ma.
\newblock Novor: real-time peptide de novo sequencing software.
\newblock {\em Journal of the American Society for Mass Spectrometry},
  26(11):1885--1894, 2015.

\bibitem{muth2018evaluating}
Thilo Muth and Bernhard~Y Renard.
\newblock Evaluating de novo sequencing in proteomics: already an accurate
  alternative to database-driven peptide identification?
\newblock {\em Briefings in bioinformatics}, 19(5):954--970, 2018.

\bibitem{mohammed2018visualizing}
Yassene Mohammed and Magnus Palmblad.
\newblock Visualizing and comparing results of different peptide identification
  methods.
\newblock {\em Briefings in bioinformatics}, 19(2):210--218, 2018.

\bibitem{cai2020itp}
Lijun Cai, Li~Wang, Xiangzheng Fu, Chenxing Xia, Xiangxiang Zeng, and Quan Zou.
\newblock Itp-pred: an interpretable method for predicting, therapeutic
  peptides with fused features low-dimension representation.
\newblock {\em Briefings in Bioinformatics}, 2020.

\bibitem{shi2021deep}
Qiang Shi, Weiya Chen, Siqi Huang, Yan Wang, and Zhidong Xue.
\newblock Deep learning for mining protein data.
\newblock {\em Briefings in bioinformatics}, 22(1):194--218, 2021.

\bibitem{tran2017novo}
Ngoc~Hieu Tran, Xianglilan Zhang, Lei Xin, Baozhen Shan, and Ming Li.
\newblock De novo peptide sequencing by deep learning.
\newblock {\em Proceedings of the National Academy of Sciences},
  114(31):8247--8252, 2017.

\bibitem{zhou2017pdeep}
Xie-Xuan Zhou, Wen-Feng Zeng, Hao Chi, Chunjie Luo, Chao Liu, Jianfeng Zhan,
  Si-Min He, and Zhifei Zhang.
\newblock pdeep: predicting ms/ms spectra of peptides with deep learning.
\newblock {\em Analytical chemistry}, 89(23):12690--12697, 2017.

\bibitem{qiao2019DeepNovoV2}
Rui Qiao, Ngoc~Hieu Tran, Lei Xin, Baozhen Shan, Ming Li, and Ali Ghodsi.
\newblock Deepnovov2: Better de novo peptide sequencing with deep learning.
\newblock {\em arXiv preprint arXiv:1904.08514}, 2019.

\bibitem{qi2017pointnet}
Charles~R Qi, Hao Su, Kaichun Mo, and Leonidas~J Guibas.
\newblock Pointnet: Deep learning on point sets for 3d classification and
  segmentation.
\newblock In {\em Proceedings of the IEEE conference on computer vision and
  pattern recognition}, pages 652--660, 2017.

\bibitem{zhao2019deep}
Minghang Zhao, Shisheng Zhong, Xuyun Fu, Baoping Tang, and Michael Pecht.
\newblock Deep residual shrinkage networks for fault diagnosis.
\newblock {\em IEEE Transactions on Industrial Informatics}, 16(7):4681--4690,
  2019.

\bibitem{roepstorff1984proposal}
Peter Roepstorff and Jan Fohlman.
\newblock Proposal for a common nomenclature for sequence ions in mass spectra
  of peptides.
\newblock {\em Biomedical mass spectrometry}, 11(11):601--601, 1984.

\bibitem{vaswani2017attention}
Ashish Vaswani, Noam Shazeer, Niki Parmar, Jakob Uszkoreit, Llion Jones,
  Aidan~N Gomez, {\L}ukasz Kaiser, and Illia Polosukhin.
\newblock Attention is all you need.
\newblock In {\em Advances in neural information processing systems}, pages
  5998--6008, 2017.

\end{thebibliography}
\end{document}